\documentstyle[12pt]{article}

\title{\marginpar{\vspace{-1in}\hspace{-1in}\small KFT U{\L} 1/93}
On the $q$-deformed Hamiltonian mechanics}

\author{P.~Caban, A.~Dobrosielski,\and A.~Krajewska, Z.~Walczak\vspace{2ex}\\
University of {\L}\'od\'z\\
Department of Theoretical Physics\\
ul.~Pomorska 149/153\\
90--236 {\L}\'od\'z, Poland }

\date{January 1993}

\begin{document}

\maketitle\vfill

\begin{abstract}

We described the $q$-deformed phase space. The $q$-deformed Hamilton eqations
of motion are derived and discussed. Some simple models are considered.

\end{abstract}\vfill

\newpage

\section{Introduction}

Non-commutative geometry has attracted much attention of theoretical
physicists. It is based on the idea, that the commutative algebra  of
functions on a manifold can be replaced by an abstract non-commutative
algebra. The ideas of non-commutative geometry lead to the concept of
non-commutative or $q$-deformed physics recently realised as a number of
simple (dynamical) models with $q$-deformed phase space structure
\cite{a,c,d,e}.
We treat a non-commutative dynamics as a non-commutative differential calculus
(deformed deRham complex). In this paper we derive the $q$-deformed Hamilton
equations
of motion from Hamilton's principle of least action.

This paper is organized as follows. In Section 2 we describe the quantum
Manin's plane and the possible differential calculi connected with it.
In Section 3 we derive the $q$-deformed Hamilton equations of motion.
Finally in Section 4 we discuss the equations of motion and consider some
simple models.

\section{Mathematical preliminaries}

The  Manin's plane is defined as the quotient algebra
${\bf M}^2_q=C(x,p)/I(xp-qpx)$,  where $C(x,p)$ is an associative
algebra with unit element over $\bf{C}$ freely generated by
$x$ and $p$ while $I$  is the two-side ideal in $C$ spanned by  monomials
containing expressions $(xp-qpx)^k$, $k > 0$. The non-commutative  phase space
is obtained by introducing in the ${\bf M}^2_q$ the ${\ast}$
antilinear  anti-involution by the relations
\begin{equation} \label{1}
 x^{\ast} = x ,\quad p^{\ast} = p.
\end{equation}
The basic reordering rule reads:
\begin{equation}          \label{2}
x p = q p x,
\end{equation}
where the hermicity requires $|q| = 1$ (i.e.\ $q$ lies on unit circle).
Now, the algebra generated by $x$ and $p$ is the zero-form  sector  of
the twisted deRham complex, generated by $x$ and $p$ and differentials
$dx$, $dp$. As was shown in \cite{b} there are three families  of  possible
differential calculi related to the Manin's plane; if we take  into
account eqs.(\ref{1},\ref{2}) the basic reordering rules read:

\subsubsection*{Family I:}
\begin{eqnarray}
x\, dx & = & r\, dx\, x\label{3}\\
p\, dp & = & s\, dp\, p \nonumber\\
x\, dp & = & q\, dp\, x \nonumber\\
p\, dx & = & q^{-1}\, dx\, p \nonumber\\
dx\, dp & = & -q\, dp\, dx \nonumber\\
(dx)^2 & = & 0 \nonumber\\
(dp)^2 & = & 0. \nonumber
\end{eqnarray}

\subsubsection*{Family II:}
\begin{eqnarray}
x\, dx & = & r\, dx\, x\label{4}\\
p\, dp & = & r\, dp\, p \nonumber\\
p\, dx & = & rq^{-1}\, dx\, p\nonumber \\
x\, dp & = & (r-1)\, dx\, p + q\, dp\, x\nonumber \\
dx\, dp & = & -q r^{-1}\, dp\, dx \nonumber\\
(dx)^2 & = & 0 \nonumber\\
(dp)^2 & = & 0.\nonumber
\end{eqnarray}

\subsubsection*{Family IIa {\rm (exceptional)}:}
\begin{eqnarray}
x\, dx & = & r\, dx\, x\label{4a}\\
p\,dp & = & -dp\, p \nonumber\\
p\, dx & = & -q^{-1}\, dx\, p \nonumber\\
x\, dp & = & -2\, dx\, p + q\, dp\, x \nonumber\\
dx\, dp & = & q\, dp\, dx \nonumber\\
(dx)^2 & = & 0 \nonumber\\
(dp)^2 & = & 0.\nonumber
\end{eqnarray}
In the eqs.(\ref{3}--\ref{4a}) $|r| = |s| = 1$.

Now, according to the paper \cite{d}  let  us  introduce  the  additional
hermitean generators $K$ and $\Lambda$:
\begin{equation}\label{5}
K^{\ast} = K ,\quad {\Lambda}^{\ast} = {\Lambda},
\end{equation}
which satisfy the following reordering rules:
\begin{eqnarray}
x {\Lambda} & = & {\xi} {\Lambda} x\label{6}\\
p {\Lambda} & = & {\eta} {\Lambda} p \nonumber\\
x K & = & {\tau} K x \nonumber\\
p K & = & {\epsilon} K p \nonumber\\
{\Lambda} K & = & {\kappa} K {\Lambda}\nonumber
\end{eqnarray}

with $|\xi|=|\eta|=|\tau|=|\epsilon|=|\kappa|=1$. The generators $K$ and
$\Lambda$ are related to the mass and $\hbar I$ respectively (for details see
\cite{d}).
So the algebra ${\bf M}^4(x,p,K,{\Lambda})$ can be abstractly  defined
as a free algebra $A(x,p,K,{\Lambda})$ divided by  two-side  ideal
$J$ determined  by  the rules (\ref{2}) and (\ref{6}), i.e.\ ${\bf M}^4=A/J$.
Moreover we also assume that the  differentials  $d{\Lambda}$  and
$dK$ vanishes:
\begin{equation}\label{7}
d{\Lambda} = dK = 0
\end{equation}
(i.e.\ $K$ and $\Lambda$ are constants of motion).
By means of the Leibniz rule it is easy to  obtain  the  remaining
reordering rules between ${\Lambda}$, $K$ and $dx$, $dp$:
\begin{eqnarray}
dx\, {\Lambda} & = & {\xi} {\Lambda}\, dx\label{8}\\
dp\, {\Lambda} & = & {\eta} {\Lambda}\, dp \nonumber\\
dx\, K & = & {\tau} K\, dx \nonumber\\
dp\, K & = & {\epsilon} K\, dp. \nonumber
\end{eqnarray}

\section{Non-commutative dynamics}

To  describe  the  motion  of  a  particle   in   terms   of   the
non-commutative coordinates we assume dependence of $x$ and $p$ on  the
trajectory parameter $t$ (time) i.e.:
\begin{equation}\label{9}
x = x(t)
\end{equation}
\[p = p(t).\]
The time derivative will be denoted by the  dot.  We  assume  that
$\dot{x}$ and $\dot{p}$ are standard (non-quantum) derivatives
of $x$ and $p$. It  means that the differential $dt$ commutes with
even differential forms  and anticommutes with odd forms.
In the following we will restrict ourselves to  the  Family  I  of
differential calculi.

To describe dynamics of a physical system  in  the framework
of  non-commutative
phase space  we accept Hamilton's principle. We start with
the action of the following form:
\begin{equation}\label{10}
S_{12}  =  \int_{t_1}^{t_2} \, dt\,  \left(  \frac{1}{2}  (\dot{x}p   +   p
\dot{x} )  - H(x,p,K,{\Lambda}) \right),
\end{equation}
where the Hamiltonian $H$ is a hermitean element  of  the  algebra
i.e.:
\begin{equation}\label{11}
H^{\ast} = H.
\end{equation}
Now the  physical  trajectories  are  followed from the
variational principle:
\begin{equation}\label{12}
{\delta} S_{12} = 0
\end{equation}
with
${\delta}x(t_1)=0$, ${\delta}x(t_2)=0$, ${\delta}p(t_1)=0$, ${\delta}
p(t_2)=0$,  where
${\delta}p$, ${\delta}x$, ${\delta}H$
 are elements of the algebra.  We  can write ${\delta}H$ in the following form:
\begin{equation}\label{13}
{\delta}H\equiv{\delta}x\,\partial_{x}^{L}H+{\delta}p\,\partial_{p}^L
H\equiv\partial_{x}^{R}H\,{\delta}x+\partial_{p}^{R}H\,{\delta}p,
\end{equation}
because $H=H(x,p,K,\Lambda )$.
Here the superscripts $L$ and $R$ denote left and right derivatives
respectively.
We can find  the   explicit   form   of    $\partial_{x}^{L}H$,
$\partial_{x}^{R}H$, $\partial_{p}^{L}H$, $\partial_{p}^{R}H$
from (\ref{13})  if  we  know  the  commutation  relations  between  the
variations ${\delta}x$, ${\delta}p$ and generators  of  the
algebra ${\bf M}^4$.
It is easy to see that:
\begin{eqnarray}
{\delta}p\,x & = & q^{-1}\,x\,{\delta}p\label{14}\\
{\delta}x\,p & = & q\,p\,{\delta}x.\nonumber
\end{eqnarray}
Moreover we postulate the following commutation relations  in  the
Bethe Ansatz form:
\begin{equation}\label{15}
x\,{\delta}x=R\,{\delta}x\,x
\end{equation}
\[p\,{\delta}p=S\,{\delta}p\,p\]
with $|R|=|S|=1$.
The other relations have not an effect on the dynamics (the  form
of equations of motion is not influenced by them).
Now, explicit form  of  $\partial_{x}^{L}H$,  $\partial_{x}^{R}H$,
$\partial_{p}^{L}H$, $\partial_{p}^{R}H$, is:
\begin{eqnarray}\label{16}
\partial_{x}^{L}H & = & \frac{1}{x}\frac{H(Rx,p,K,{\Lambda})-H(x,p,K,\Lambda)}
{R-1}\\
\label{17}
\partial_{p}^{L}H & = & \frac{1}{p}\frac{H(x,Sp,K,{\Lambda})-H(x,p,K,
\Lambda)}{S-1}\\
\label{18}
\partial_{x}^{R}H & = & \frac{1}{x}\frac{H(R^{-1}x,qp,\tau K,\xi
\Lambda )-H(x,qp,\tau K,\xi \Lambda )}{R^{-1}-1}\\
\label{19}
\partial_{p}^{R}H & = & \frac{1}{p}\frac{H(q^{-1}x,Sp,\epsilon K,\eta\Lambda )
-H(q^{-1}x,p,\epsilon K,\eta\Lambda )}{S^{-1}-1}
\end{eqnarray}
We see that (\ref{16}) and (\ref{17}) are the Gauss-Jackson derivatives, while
(\ref{18}) and (\ref{19}) are the rescaled ones. Putting $\delta$H from
eqs.(\ref{13},\ref{16}--\ref{19})    to    eq.(\ref{12})    and    integrating
(\ref{12}) by parts we  obtain  the following equations of motion:
\begin{eqnarray}\label{20}
\frac{1+q}{2}\dot{x} & = & -\frac{1}{p}\frac{H(x,Sp,K,\Lambda )-H (x,p,K
,\Lambda )}{S-1}\\
\frac{1+q}{2}\dot{p} & = & \frac{1}{x}\frac{H(R^{-1}x,qp,\tau K,\xi\Lambda )
-H(x,qp,\tau K,\xi\Lambda )}{R^{-1}-1}\nonumber
\end{eqnarray}
Consistency of these equations with algebra ${\bf M}^4$  gives the following
conditions:
\begin{eqnarray}\label{21}
q^{-1}H(x,qp,\tau K,\xi\Lambda ) & = & H(x,p,K,\Lambda )\\
qH(q^{-1}x,p,\epsilon K,\eta\Lambda ) & = & H(x,p,K,\Lambda )\nonumber\\
\epsilon\tau H(\tau^{-1}x,\epsilon^{-1}p,K,\kappa^{-1}\Lambda ) &
= & H(x,p,K,\Lambda )\nonumber\\
\eta\xi H(\xi^{-1}x,\eta^{-1}p,\kappa K,\Lambda ) & = & H(x,p,K,\Lambda
)\nonumb
er\\
r =s= 1\nonumber
\end{eqnarray}
(Note that every condition above we  can  derive  from  the  eq.(\ref{10})
demanding that $H$ must commute with every generator of  the  algebra
in the same way as $(dx)p$).

Because  the Hamiltonian is also an element of  the  algebra  we  can
write it as a formal power series of generators:
\begin{equation}\label{22}
H(x,p,K,\Lambda )=\sum_{n,m,k,l}c_{nmkl}x^{n}p^{m}K^{k}\Lambda^{l}.
\end{equation}
Hermicity gives additional conditons:
\begin{equation}\label{23}
R^{n}=1\quad {\rm or}\quad R^{n-1}=1
\end{equation}
\[S^{m}=1\quad {\rm or}\quad S^{m-1}=1.\]

\section{Motion of a particle}

Let us choose the Hamiltonian in the form:
\begin{equation}\label{24}
H=\epsilon^{-2}p^{2}K^{2}+V(x,K,\Lambda ),
\end{equation}
where  the $\epsilon^{-2}p^{2}K^{2}$   is  the  kinetic  term  and
$V(x,K,\Lambda )$  is   a potential. We demand:
\begin{equation}\label{25}
V^{\ast}(x,K,\Lambda )=V(x,K,\Lambda ).
\end{equation}
Equations of motion are the following:
\begin{eqnarray}\label{26}
\frac{1+q}{2}\dot{x} & = & -(S+1)\epsilon^{-2}pK^{2}\\
\frac{1+q}{2}\dot{p} & = & \frac{1}{x}\frac{V(R^{-1}x,\tau K,\xi\Lambda )-
V(x,\tau K,\xi\Lambda )}{R^{-1}-1}.\nonumber
\end{eqnarray}
{}From eqs.(\ref{23}) and by means of the classical limit existence we
have:
\begin{equation}\label{27}S=1.\end{equation}
The following equations arise from conditions (\ref{21}) (again under the
assumption of the existence of the proper classical limit):
\begin{eqnarray}
\epsilon & = & q^{-1/2}\nonumber\\
\tau & = & q^{-1/2}\label{28}\\
\xi\kappa^{2} & = & \eta\nonumber
\end{eqnarray}
so finally:
\begin{eqnarray}\label{29}
qV(q^{-1}x,q^{-1/2}K,\eta\Lambda ) & = & V(x,K,\Lambda )\\
q^{-1}V(x,q^{-1/2}K,\eta\kappa^{-2}\Lambda ) & = & V(x,K,\Lambda )\nonumber\\
\eta^2\kappa^{-2}V(\eta^{-1}\kappa^2x,\kappa K,\Lambda ) & = & V(x,K,\Lambda
)\n
onumber\\
q^{-1}V(q^{1/2}x,K,\kappa^{-1}\Lambda ) & = & V(x,K,\Lambda ).\nonumber
\end{eqnarray}
Now we restrict our potential to the form
\begin{equation}\label{30}
 V(x,K,\Lambda )=f(x^{N}K^{M}\Lambda^{L})K^{A}\Lambda^{B}.
\end{equation}
Then the relations (\ref{29}) can be rewritten:
\begin{eqnarray}\label{31}
qf(q^{-\frac{2N+M}{2}}\eta^{L}x^{N}K^{M}\Lambda^{L})q^{\frac{2B-A}
{2}} & = & f(x^{N}K^{M}\Lambda^{L})\\
q^{-1}f(q^{-\frac{M}{2}}\eta^{L}\kappa^{-2L}x^{N}K^{M}\Lambda^{L})
q^{-\frac{A}{2}}\eta^{B}\kappa^{-2B} & = & f(x^{N}K^{M}\Lambda^{L})\nonumber\\
\eta^{2}\kappa^{-2}f(\eta^{-N}\kappa^{2N+M}x^{N}K^{M}\Lambda^{L})\kappa^{A
} & = & f(x^{N}K^{M}\Lambda^{L})\nonumber\\
q^{-1}f(q^{\frac{N}{2}}\kappa^{-2}x^{N}K^{M}\Lambda^{L})\kappa^{-B}
& = & f(x^{N}K^{M}\Lambda^{L})\nonumber
\end{eqnarray}
Introducing:
\begin{equation}\label{32}
Y=x^{N}K^{M}\Lambda^{L}
\end{equation}
we can rewrite conditions (\ref{31}) as:
\begin{equation}\label{33}
f(q^{\frac{-2NB-MB+L(A-2)}{2B}}\lambda_{1}^{-\frac{L}{B}}Y)=\lambda_{1}f(Y)
\end{equation}
\[\lambda_{1}=q^{\frac{A-2}{2}}\eta^{-B}\]\vspace{\baselineskip}
\begin{equation}\label{34}
f(q^{\frac{-MB+L(A+2)}{2B}}\lambda_{2}^{-\frac{L}{B}}Y)=\lambda_{2}
f(Y)
\end{equation}
\[\lambda_{2}=q^{\frac{A+2}{2}}\eta^{-B}\kappa^{2B}\]\vspace{\baselineskip}
\begin{equation}\label{35}
f(\kappa^{\frac{-N(2-A)+4N+2M}{2}}\lambda_{3}^{\frac{N}{2}}Y)=\lambda_{3}
f(Y)
\end{equation}
\[\lambda_{3}=\kappa^{2-A}\eta^{-2}\]\vspace{\baselineskip}
\begin{equation}\label{36}
f(\kappa^{-\frac{NB+2L}{2}}\lambda_{4}^{\frac{N}{B}}Y)=\lambda_{4}f
(Y)
\end{equation}
\[\lambda_{4}=q\kappa^{B}.\]
It is natural to assume that all of above relations (\ref{33}--\ref{36}) are
derived from the only one scaling relation on the function $f$:
\begin{equation}\label{37}
f(\lambda^{b}Y)=\lambda f(Y).
\end{equation}
We also assume that $q$ is not  necessary  a  root  of  unity. So we
have:
\begin{eqnarray}\label{38}
2M+N(A+2) & = & 0\\
NB+2L & = & 0\nonumber\\
\lambda_{1}=\lambda_{2}=\lambda_{3}=\lambda_{4}=1.\nonumber
\end{eqnarray}
Therefore:
\begin{eqnarray}
\eta & = & q^{\frac{A-2}{2B}}\label{39}\\
\kappa & = & q^{-\frac{1}{B}}\nonumber
\end{eqnarray}
and from (\ref{28}) we get
\begin{equation}\label{40}
\xi =q^{\frac{A+2}{2B}}
\end{equation}
uniqueness of the above $B$-th degree root is guaranted by the requirement of
the existnce of classical limit i.e.\ when $q\rightarrow 1$ then
$\eta\rightarrow 1$, $\kappa\rightarrow 1$ and $\xi\rightarrow 1$.
Let us denote:
\begin{equation}\label{41}
{\bf X}=q^{-\frac{A+2}{8}}xK^{-\frac{A+2}{2}}\Lambda^{-\frac{B}{2}}.
\end{equation}
Then equations of motion take the form:
\begin{eqnarray}\label{42}
\dot{x} & = & -\frac{4q}{1+q}pK^{2}\\
\dot{p} & = & \frac{2q}{1+q}\frac{1}{x}\frac{f(R^{-N}\alpha{\bf X}^{N})-f
(\alpha{\bf X}^{N})}{R^{-1}-1}K^{A}\Lambda^{B},\nonumber
\end{eqnarray}
where the phase-factor $\alpha$ can be derived from:
\begin{equation}\label{43}
\alpha {\bf X}^{N}=x^{N}K^{-\frac{(A+2)N}{2}}\Lambda^{-\frac{BN}{2}}.
\end{equation}
Note that $\bf X$ is hermitean and belongs  to  the  center  of  algebra
${\bf M}^{4}$.

Now, let us consider the dynamical models described by Aref'eva and  Volovich
\cite{a}.

\subsubsection*{Free particle}

We choose the potential $V=0$ so $H=qp^{2}K^{2}$ and consequently
\begin{eqnarray}\label{44}
\dot{x} & = & -\frac{4q}{1+q}pK^{2}\\
\dot{p} & = & 0.\nonumber
\end{eqnarray}
Notice that eqs.(\ref{44}) do not contain $\Lambda$.

\subsubsection*{Harmonic oscillator}

We start with Hamiltonian:
\begin{equation}\label{45}
H=qp^{2}K^{2}+x^{2}K^{-2-A}\Lambda^{-B}K^{A}\Lambda^{B}.
\end{equation}
To obtain (\ref{45}) we choose $f$ the identity and $N=2$ in (\ref{30}).
\begin{eqnarray}\label{46}
\dot{x} & = & -\frac{4q}{1+q}pK^{2}\\
\dot{p} & = & \frac{4q}{1+q}q^{A}xK^{-2}.\nonumber
\end{eqnarray}\vspace{2\baselineskip}

Now, we turn back to the general case.

Let us introduce another hermitean operator ${\bf P}$:
\begin{equation}\label{47}
{\bf P}=q^{\frac{A-2}{8}}p\Lambda^{-\frac{B}{2}}K^{-\frac{A-2}{2}}
\end{equation}
It is easy to verify , that eqs.(\ref{42})  can  be  rewritten  in  the
form:
\begin{eqnarray}\label{48}
\dot{\bf X} & = & -\frac{4q}{1+q}q^{-1/2}{\bf P}\\
\dot{\bf P} & = &\frac{2q}{1+q}q^{\frac{A-1}{2}}\frac{1}{{\bf
X}}\frac{f(R^{-N}\alpha      {\bf      X}^{N})-f(\alpha      {\bf
X}^{N})}{R^{-1}-1}\nonumber
\end{eqnarray}
and we have:
\begin{equation}     \label{49}
{\bf XP}={\bf PX}.
\end{equation}

\section{Conclusions}

We  described a  $q$-deformation of the  classical   Hamiltonian mechanics.
To derive the $q$-deformed Hamilton equations of motion we deformed the phase
sp
ace leaving the principle of least action unchanged.
We were able to reduce number of free deformation parameters of theory
requiring
the existence of a smooth classical limit. In this last case we found a
transfor
mation of variables which
allowed us to replace deformed commutation relations by the standard ones.

\section*{Acknowledgments}
We are grateful for interesting discussions and helpful comments\linebreak
to Prof.~J.~Rembieli\'nski and Mr.~K.A.~Smoli\'nski.

This work is supported by KBN grant No.\ 2 0218 91 01.

\end{document}